# The feasibility of artificial consciousness through the lens of neuroscience

Jaan Aru[1], Matthew E. Larkum[2] & James M. Shine[3]

[1]Institute of Computer Science, University of Tartu, Estonia
[2]Institute of Biology, Humboldt University Berlin, Germany
[3]Brain and Mind Center, The University of Sydney, Sydney, Australia

correspondence:
jaan.aru@ut.ee (J.Aru); larkumma@hu-berlin.de (M.Larkum), mac.shine@sydney.edu.au (J.M.Shine)

**Abstract**

Interactions with large language models have led to the suggestion that these models may soon be conscious. From the perspective of neuroscience, this position is difficult to defend. For one, the inputs to large language models lack the embodied, embedded information content characteristic of our sensory contact with the world around us. Secondly, the architecture of large language models is missing key features of the thalamocortical system that have been linked to conscious awareness in mammals. Finally, the evolutionary and developmental trajectories that led to the emergence of living conscious organisms arguably have no parallels in artificial systems as envisioned today. The existence of living organisms depends on their actions, and their survival is intricately linked to multi-level cellular, inter-cellular, and organismal processes culminating in agency and consciousness.

Keywords: artificial intelligence; consciousness; large language models; thalamus;

**Large language models and consciousness**

There is a long tradition of research asking which animals are conscious [1-3] and whether entities outside the animal kingdom might be conscious [4-6]. Recently, the advent of Large Language Models (LLMs) has brought a novel set of perspectives to this question. Through their competence and ability to converse with us, which in humans is indicative of being conscious, LLMs have forced us to refine our understanding of what it means to understand, to have agency, and to be conscious.

LLMs are sophisticated, multi-layer artificial neural networks whose weights are trained on hundreds of billions of words from various texts including natural language conversations between awake, aware humans. Through text-based queries, users interacting with LLMs are provided with a fascinating language-based simulation. If you take the time to use these systems, it is hard not to be swayed by the apparent depth and quality of the internal machinations in the network. Ask it a question, and it will provide you with an answer that drips with the kinds of nuance we typically associate with conscious thought. As a discerning, conscious agent yourself, it's tempting to conclude that the genesis of the response arose from a similarly conscious being – one that thinks, feels, reasons and experiences. Using this type of a "Turing test" as a benchmark, the question can be raised whether LLMs are or soon will be conscious [7-10], which in turn raises a host of moral quandaries, such as whether it is ethical to continue to develop LLMs that could be on the precipice of conscious awareness. While this position might not be prevalent among neuroscience researchers today, the improving capabilities of AI systems will inevitably lead to the point where the possibility of machine consciousness needs to be addressed. Furthermore, this possibility is discussed extensively in media and thus neuroscientists need to consider some of the arguments in favor and against it.

This perspective is often bolstered by the fact that the architecture of LLMs is loosely inspired by features of brains (Fig. 1) – the only objects to which we can currently (and confidently) attribute consciousness. However, while early generations of artificial neural networks were designed as a simplified version of the cerebral cortex [11], modern LLMs have been highly engineered and fit to purpose in ways that do not retain deep homology with the known structure of the brain. Indeed, many of the circuit features that render LLMs computationally powerful (Fig. 1) have strikingly different architectures from the systems to which we currently ascribe causal power in the production and shaping of consciousness in mammals. For instance, many theories of the neural basis of consciousness would assign a central role in conscious processing to thalamocortical [12–17] and arousal systems [18–24], both features that are architecturally lacking in LLMs.

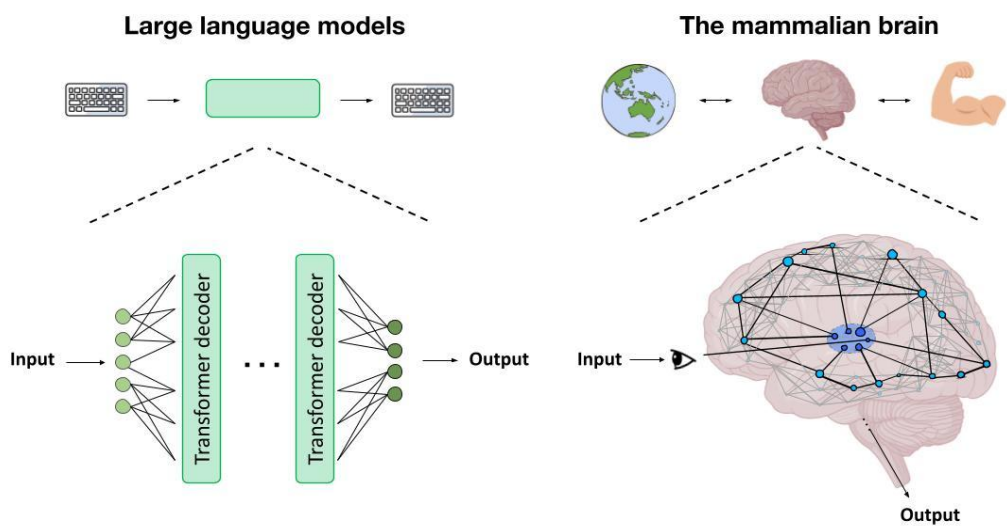

**Figure 1 – Macroscopic topological differences between mammalian brains and large language models.** Left – a schematic depicting the basic architecture of a large language model, which can have tens or even more than a hundred decoder blocks arranged in a feed-forward fashion. Right – a heuristic map of the thalamocortical system, which generates complex activity patterns thought to underlie consciousness.

One might ask why it is so crucial for the architecture of LLMs to mimic features of the brain. The primary reason is that the only version of consciousness that we can currently be absolutely sure of arises from brains embedded within complex bodies. This could be further collapsed to humans, though many of the systems-level features thought to be important for subjective consciousness are pervasive across phylogeny, stretching back to mammals [13,24,25], and even to invertebrates [26]. We will return to this point, but we will start with the question about precisely what we mean by the term 'consciousness'. From there we will present the three arguments against the view that present-day AI systems have, or that future AI systems will soon have, consciousness. First, consciousness is tied to the sensory streams that are meaningful for the organism; second, in mammalian brains, consciousness is supported by a highly interconnected thalamocortical system that is strikingly different from the architecture of LLMs; and third, that consciousness might be inextricably linked to the complex biological organization characteristic of living systems.

**What is consciousness?**

Typically, people rely on interactive language-based conversations to develop an intuitive sense about whether LLMs are conscious or not. Although these conversations are remarkable, they are not formal objective measures of consciousness and do not constitute *prima facie* evidence for conscious agency. The advent of LLMs has demanded a re-evaluation of whether we can indeed infer consciousness directly from interactions with other agents. Thus, there is an emerging view that the criteria for attributing human-like intelligence need to be re-assessed [27]. To make sensible progress on this matter, we need to better understand what exactly people think and assume when talking about consciousness.

There are different meanings associated with the word "consciousness": neurologists often refer to levels of consciousness (e.g., the fact that you are [or are not] conscious; i.e., state of consciousness), whereas psychologists often interrogate the contents of consciousness (e.g., consciousness *of* something; i.e., content of consciousness) [17]. Furthermore, there is a distinction between different contents of consciousness: our experiences can be described as primarily phenomenal [28] (e.g., experiential, the sight/smell of an apple; or the feel of your arm) or more cognitive (i.e., how we access and report conscious experiences [28]; or how we manipulate abstract ideas and concepts, such as our sense of self, or ephemeral ideas such as "justice" or "freedom"). The question of whether AI systems are conscious could include only one or all of these aspects. Here, we mainly focus on phenomenal consciousness and ask whether machines can experience the world phenomenally.

**The Umwelt of an LLM**

The portion of the world that is perceptually 'available' to an organism has been described as its *"Umwelt"* (from the German 'environment' [29]). For instance, human retinas respond to wavelengths of light ranging from ~380-740 nanometers, which we perceive as a spectrum from blue to red. Without technological augmentation, we are not susceptible to light waves outside of this narrow band – in the infrared (>740 nm) or ultraviolet (<380 nm) bands. We have a similar Umwelt for the auditory (we can perceive tones between 20-20,000 Hz), somatosensory (we can differentiate stimulation up to 1 mm apart on some parts of our body) and vestibular domains (yoked to the 3-dimensional structure of our semicircular canals). Other species can detect other portions of the electromagnetic spectrum. For instance, honeybees can see light in the ultraviolet range [30], and some snakes can detect infrared radiation in addition to more traditional visual light cues [31] – that is, the bodies and brains of other animals place different constraints on their susceptibility to the sensory world around them. James Gibson referred to this information – that is pragmatically *available* to us – as a set of "affordances" [25,32-34].

If anything at all, what is the Umwelt of an LLM? What kinds of affordances does an LLM have access to? By the nature of its design, an LLM is only ever presented with binary-coded patterns fed to the network algorithms inherent within the complex transformer architectures that comprise the inner workings of present-day LLMs [35,36]. While neuronal spikes also potentially encode incoming analog signals as digital (i.e., binary), the information stream fed to the LLMs in question is highly abstract, and hence does not itself make robust contact with the world *as it is* in any way. Text and strings of keystrokes on a keyboard are simply no match for the dynamic complexity of the natural world: the Umwelt of an LLM – the information afforded to it – is written in a completely different language to the information that bombards our brain from the moment we open our eyes. In this regard, the information stream provided to LLMs is more different from the one presented to humans than ours is from bats [37].

With all this said, it bears mention that there is nothing stopping the input of future AI systems from being much more enriched – i.e., LLMs could be equipped with different types of inputs (see [38,39]) that better match the statistics of the natural world than the small slice of electromagnetic signals available to human brains. Could the Umwelt of future AI systems become more extended than that available to humans? It is essential to recognize that our Umwelt and conscious experience are not determined solely by sensory input. For example, consider lying in a floating tank where, despite a lack of normal sensory experiences, consciousness does not extinguish. This underscores the notion that having an Umwelt presupposes an inherent subjective perspective – that is, an agent to begin with [29,40,41]. Similarly, affordances depend also on the internal properties of the agents, in particular their motivations and goals [33,40,41]. This underscores the point that consciousness does not arise merely from data, and hence that simply adding massive data streams to future AI systems will not lead to consciousness.

This perspective allows us to rethink some of our fundamental assumptions in the science of consciousness. Specifically, as AI systems continue to exhibit increasingly sophisticated abilities, it prompts us to re-evaluate the necessity of more basic self- and agency-related processes for the emergence of consciousness. Although there are theories of consciousness that make this assumption [42-46], many theories take the classic information processing view where consciousness 'pops out' at some elaborated stage of stimulus processing. The contrast between biological consciousness and LLMs demonstrates that we must reconsider our assumptions. Perhaps, to be conscious, the external world must be integrated with the internal needs and processes.

**The neural architecture supporting conscious integration**

There is a sizable literature on the neural correlates of consciousness, with many different theories about the neural processes that underlie conscious processing [47]. Many of these frameworks highlight that consciousness is supported by neural processing within the dense, re-entrant thalamocortical network [12-17,48-53]. The thalamocortical network encompasses cortical areas, cortico-cortical connectivity, and higher-order thalamic nuclei with their diffuse projections to cortical areas [53-56]. This specific architecture of the thalamocortical system supports recurrent and complex processing thought to underlie consciousness [57-61] and conscious integration, i.e., the fact that despite processing happening in different areas, consciousness is unified [51,53,62]. However, the details of how this integration is achieved are different in different theories of consciousness.

For instance, according to the Global Neuronal Workspace Theory [48,49] consciousness depends on the central workspace constituted by a distributed frontoparietal cortical system. This workspace integrates information from local cortical processors and then globally broadcasts it to all local processors, with the global broadcast delineating conscious from non-conscious processes. Other theories of consciousness assign a different neural process to carry out this integration. For instance, the integrated information theory of consciousness [50,51] suggests that conscious integration mainly occurs in posterior cortical regions. The binding-by-synchrony theory [63,64], which is less influential today, suggested that conscious integration occurs via a process – high-frequency synchronization between different cortical areas – that can be putatively involved in diverse functions including perception, cognition or motor planning, depending on the cortical regions involved.

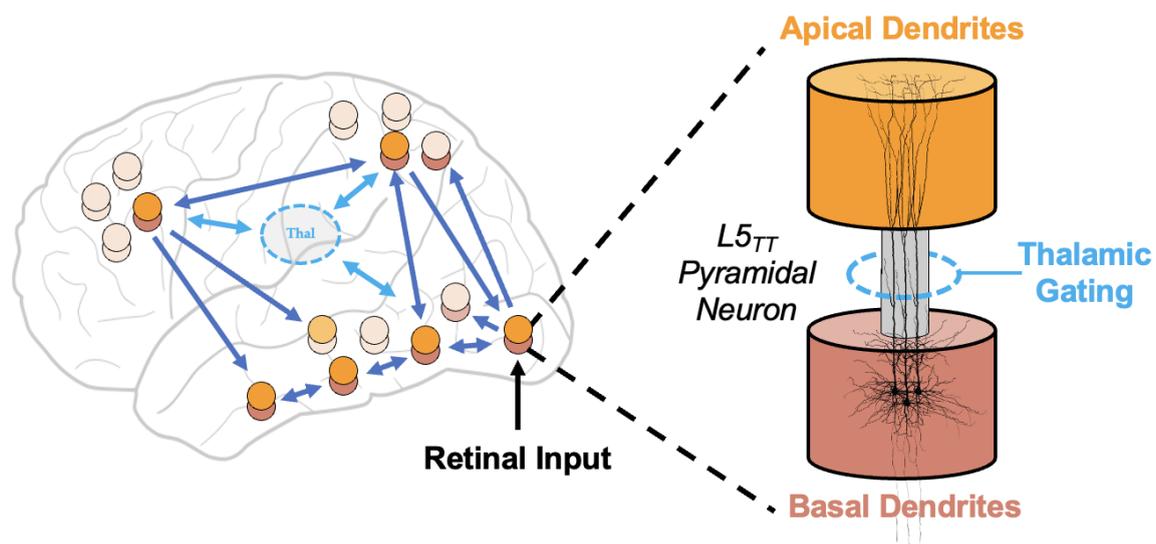

**Figure 2 – The neural architecture underlying conscious integration according to the Dendritic Integration Theory**. (left): when inputs hit the retina, they contain information content that will drive feed-forward activity in the ventral visual stream, however only the patterns that are coherent with the information content will be augmented (e.g., bilateral arrows and simultaneously active apical/basal compartments of the neurons in ventral stream). The top-down prediction of these features from frontal or parietal cortex can augment certain features of the input stream, causing them to stand out from the background, leaving others inactive (e.g., the light red basal compartment in primary visual area). According to the DIT [12,53], the thalamus (light blue; dotted line) plays a crucial role in shaping/gating the contents of consciousness. (right) DIT associates consciousness with the subset of thick-tufted layer 5 (L5tt) pyramidal neurons that are burst-firing, which occurs when depolarisation of the cell body via basal dendrites (red) coincides temporally with descending cortical inputs to apical dendrites (orange), particularly in the presence of gating inputs from higher-order, matrix-type thalamus (blue).

In Dendritic Integration Theory (DIT; [12,53]; Fig. 2), it is proposed that global conscious integration also depends on local integration on the level of single layer 5 pyramidal neurons, which are large excitatory neurons that hold a central position in both thalamocortical and corticocortical loops [12,53]. These neurons have two major compartments (Fig. 3, orange and red cylinders) that process categorically distinct types of information: the basal compartment (red) processes externally-grounded information whereas the apical compartment (orange) processes internally-generated information [12,53,65]. According to DIT, during conscious states, these two compartments are coupled, i.e., integrated, allowing information to flow through the thalamocortical and corticocortical connections, thus enabling system-wide integration and consciousness [12,53].

Notably, the architecture of present-day LLMs is devoid of features from each of these theoretical proposals: there is no LLM equivalent of dual-compartment pyramidal neurons, nor a centralised thalamic architecture, nor a global workspace, nor the many arms of the ascending arousal system. In other words, LLMs are missing the very features of brains that are currently hypothesized to support consciousness. Although we are not arguing that the mammalian brain is the *only* architecture capable of supporting conscious awareness, the evidence from neurobiology suggests that very specific architectural principles (i.e., more than simple connections between integrate-and-fire neurons) are responsible for mammalian consciousness (see [1-4, 26] for some examples of research on consciousness in non-mammalian species). According to the best evidence from nature, therefore, we are hesitant to ascribe phenomenal consciousness to present-day LLMs, which are topologically extremely simple in comparison.

Could future AI models eventually incorporate the process of integration? The integration proposed by Global Neuronal Workspace Theory [48,49] offers a relatively straightforward implementation [9,10] and, in fact, some recent AI systems have incorporated something akin to a global workspace shared by local processors [66,67]. As the computational process of global broadcasting can be implemented in AI systems, an artificial system with a computationally equivalent global workspace ought to be considered conscious by proponents of this theory [9,10]. However, as indicated above, not all other theories would agree that this type of integration is the key to consciousness. For instance, the integrated information theory of consciousness [50,51] claims that it is impossible for a software-based AI system to achieve consciousness because computer software does not have real cause-effect power necessary for integrating information [68,69]. Here we will consider the third possibility, namely that consciousness might be implementable in principle, but it might require a level of computational specificity that is beyond the present-day (and perhaps future) AI systems.

**Consciousness as a complex biological process**

Consciousness does not only depend on the architecture of the system. For instance, the structure of the thalamocortical system does not change when we are in deep sleep or undergo anesthesia, yet consciousness disappears. Even local neural responses and gamma-band activity in primary sensory areas can remain the same, even though we are unconscious [70,71]. This implies that consciousness relies on specific neural *processes* that are different in conscious and unconscious brains.

To appreciate our current knowledge about the details distinguishing conscious from unconscious processing, we will return to DIT, as it perhaps best encapsulates the specific neurobiological nuances relevant to the matter. In particular, DIT proposes that the crucial difference between conscious and unconscious processing lies in the integration between the two compartments of pyramidal cells (Fig 2). As indicated above, during conscious processing, these two compartments interact and hence enable complex and integrated processing across the thalamocortical system [12,53]. However, it has been demonstrated that various different anesthetic agents lead to functional decoupling between the two compartments [72]. In other words, while anatomically the neurons are intact and can fire action potentials, physiologically there is no dendritic integration within these cells: the top-down feedback component cannot influence processing. This dendritic coupling was demonstrated to be controlled by metabotropic receptors, which are a special type of receptor not usually modeled in artificial neural networks. Furthermore, the authors provided suggestive evidence that the activity of the metabotropic receptors might be controlled by the higher-order thalamus [72]. Thus, there are relatively specific neurobiological processes that may be responsible for switching consciousness on and off in the brain.

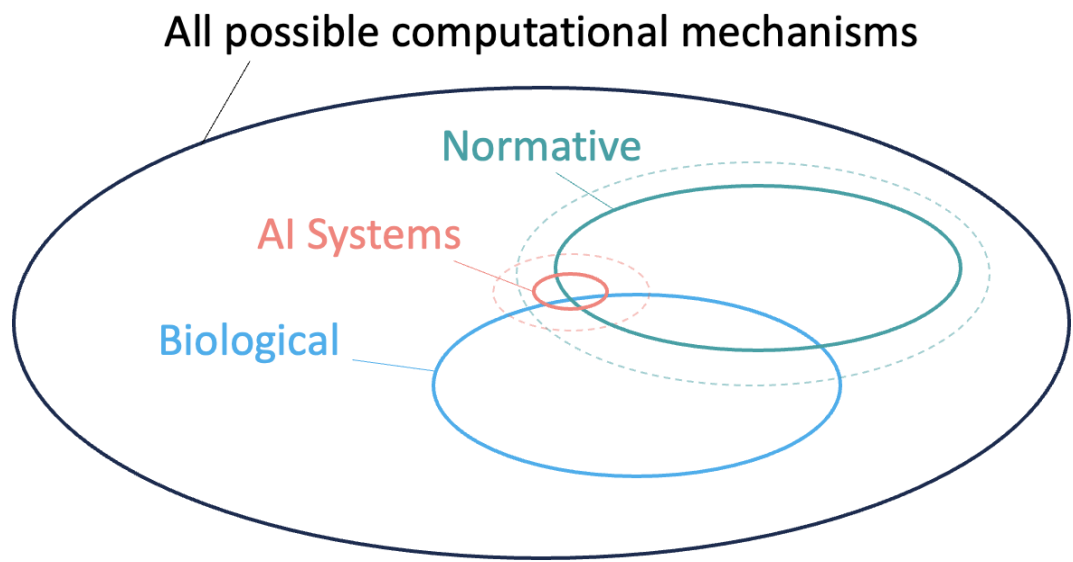

**Figure 3 - The limitations of our current computational understanding of consciousness.** The space of all possible computations (the big ellipse) is larger than the types of computations we do understand (teal ellipse), hence we might not have yet captured the key computations underlying consciousness. Although we understand some of the normative computations that comprise the biological processes responsible for consciousness (overlap between teal and blue ellipse), our knowledge horizon of computational mechanisms constantly evolves and perhaps needs to be extended to understand consciousness (dashed teal ellipse). Computations in AI systems (red ellipse) show some overlap with biological computations, and some of the computations of AI systems are understood. However, the computations of AI systems are different from those of a biological system and hence, a priori, there is little reason to think that the computations of present-day AI systems are related to computations underlying consciousness.

However compelling, these details almost certainly pale compared to the true complexity of the depth of biological organization required to achieve a satisfactory understanding of consciousness. While today's explanations of consciousness rely on ideas such as the global workspace [48,49], information integration [50,51], recurrent feedback [57,58], dendritic integration [12,53], and other notions [47] it might be the case that the biological processes underlying consciousness are more intricate than these current concepts appreciate. It is also quite possible that the abstract computational-level ideas that we currently use to frame discussions in consciousness research may miss the necessary computational details required to satisfactorily explain consciousness. In other words, biology is complex, and our understanding of biological computations is limited (Fig. 3) – so perhaps we simply do not have yet the right mathematical and experimental tools to understand consciousness.

To better ground this notion of biological complexity, it is worth considering that the cellular-level and system-level processes described above are inextricably embedded within a living organism. Living organisms differ from present-day machines and AI algorithms, as they are constantly in the process of self-maintenance across several levels of processing [73,74] (i.e., they have skin in the game (see Box 1)). While here we are not arguing that consciousness can only arise in living systems [74-77], we want to draw attention to the fact that living systems have organizational complexity – i.e., interactions between the different levels of the system [77-81] – that is not captured within present-day computer software. So while AI algorithms might model and capture neural computations happening at the level of neural circuits and large-scale networks, these algorithms do not simulate the processes within the neurons. This is a reasonable abstraction if one assumes that consciousness occurs at the level of interactions between neurons. However, the truth is that we do not know whether that assumption is correct.

A biological neuron is not just an abstract entity that can be fully captured with a few lines of code. Biological cells have multi-level organization, and depend on a further cascade of biophysical intracellular complexity [79-83]. For instance, consider the Krebs cycle that underlies cellular respiration, a key process in maintaining cellular homeostasis [84]. Cellular respiration is a crucial biological process that enables cells to convert the energy stored in organic molecules into a form of energy that can be utilized by the cell, however this process is not compressible into software as processes like cellular respiration need to happen with real physical molecules. Note that our aim is not to suggest that consciousness requires the Krebs cycle, but rather to highlight that perhaps consciousness is similar: it cannot be abstracted away from the underlying machinery [68-69,85].

Importantly, we are not claiming that consciousness cannot be captured within software at all [68-69,85-87]. Rather, we emphasize that we have to at least entertain the possibility that consciousness is linked to the complex biological organization underlying life [74-81], and thus any computational description of consciousness will be much more complex than our present-day theories suggest (Fig. 3). It might be impossible to "biopsy" consciousness and remove it from its organizational dwellings. This idea contradicts most of today's theories of consciousness which assume that consciousness can be captured on the abstract level of computation [47,88]. However, this might be another assumption about consciousness that will require updating in light of modern AI systems: perhaps interdependencies and organization across scales observed in living systems cannot be ignored if we want to understand consciousness.

It might be that although AI systems (at least to some extent) mimic their biological counterparts on the level of network computations, in these systems we have abstracted away all the other levels of processing that causally contribute to consciousness in the living brain. Hence, we have abstracted away consciousness itself. In this way, LLMs and future AI systems may be trapped in a compelling simulation of the signatures of consciousness, without any conscious experience. If consciousness is indeed related to these other levels of processing, or their coherent interactions across scales, we might still be far from conscious machines.

**Concluding remarks**

Here, we have attempted to provide a systems neuroscientists perspective on LLMs. We conclude that, while fascinating and alluring, LLMs are not conscious, and will likely not be conscious soon. Firstly, we detailed the vast differences between the Umwelt of mammals – the 'slice' of the external world that they can perceive – and the Umwelt of LLMs, with the latter being highly impoverished and limited to keystrokes, rather than the electromagnetic spectrum. Secondly, we argued that the topological architecture of LLMs, while highly sophisticated, is sufficiently different from the neurobiological details of circuits empirically linked to consciousness in mammals that there is no a priori reason to conclude that they are capable of phenomenal consciousness (Fig. 1). Third, we pointed out that it might not be possible to abstract consciousness away from the organizational complexity that is inherent within living systems but strikingly absent in AI systems. In toto, we believe that these three arguments make it extremely unlikely that LLMs, in their current form, have the capacity for phenomenal consciousness. Rather, they mimic signatures of consciousness that are implicitly embedded within the language that humans use to describe the richness of our conscious experience.

Rather than representing a deflationary account, we foresee several useful implications from this perspective. For one, we should worry much less about any potential moral quandary regarding sentience in LLMs **(Box 1)**. In addition, we believe that a refined understanding of the similarities and differences in the topological architecture of LLMs and mature brains provides opportunities for advancing progress in both machine learning (by mimicking features of brain organisation) and neuroscience (by learning how simple distributed systems can process elaborate information streams) [89-91]. For these reasons, we are optimistic that future collaborative efforts between AI and systems neuroscience have the potential to rapidly improve our understanding of how our brains make us conscious.

**Box 1: LLMs and skin-in-the-game - do we have a moral quandary?**

Does it really matter if LLMs are conscious? If LLMs can match and even exceed our expectations in terms of getting superficially human-like responses that are useful and informative, is there any need to speculate about what an LLM experiences? Some argue forcefully, yes from a moral perspective [92]. According to this view, we should carefully consider the ethical implications for any conscious entity, including artificial intelligence, principally because it is assumed that some experiences could be negative and that in this case the AI could also suffer. At this point, it is claimed, we should care or at least be cautious about whether an AI really does suffer.

To this extent, we predict that LLMs do not (and will not) have experiences that can be considered suffering in any sense that should matter to human society. One way of arguing this point is analogous to the notion of 'skin in the game' [93] which emphasizes the importance of personal investment and engagement in moral decision-making, and suggests that those who have a personal stake in an issue are more competent to make ethical judgments than those who do not [93]. An LLM could, in principle, claim in a conversation that it does not want to be shut down, but an LLM does not have *"skin in the game"*, as there is no real consequence to the software when it is actually shut down. In contrast, in biology the system has something to lose on several levels [73]: it cannot stop living, as otherwise it will die. As the philosopher Hans Jonas has said: *"The organism has to keep going, because to be* *going is its very existence"* [94]. If cellular respiration stops, the cell dies; if cells die, organs fail; if organs fail, the organism will soon die. The system has skin in the game across levels of processing [73] and these levels are not independent from each other. There is complex causality between real physical events at the microscale and consciousness. Here, we would argue that not having the capacity for phenomenal consciousness would preclude suffering and therefore personal investment. This reasoning also extends to the common disincentives that are used for legal issues that LLMs could become entangled with such as contracts, libel, etc., which are commonly penalized with actions ranging from monetary compensation to incarceration. Without personal investment on the part of the LLM, these disincentives will not be taken seriously by injured parties and would therefore likely destabilize the rule of law.

**Acknowledgements**

We would like to thank Jakob Howhy, Kadi Tulver, Raul Vicente, Christopher Whyte and Gabriel Wainstein for their helpful comments on the manuscript. This research was supported by the European Social Fund through the "ICT programme" measure, by the European Regional Development Fund through the Estonian Centre of Excellence in IT (EXCITE), and the Estonian Research Council grant PSG728.

**Declaration of interests**

The authors declare no conflict of interest

# References

1. Seth, A. K., Baars, B. J., & Edelman, D. B. (2005). Criteria for consciousness in humans and other mammals. Consciousness and cognition, 14(1), 119-139.
2. Edelman, D. B., & Seth, A. K. (2009). Animal consciousness: a synthetic approach. Trends in neurosciences, 32(9), 476-484.
3. Birch, J., Schnell, A. K., & Clayton, N. S. (2020). Dimensions of animal consciousness. Trends in cognitive sciences, 24(10), 789-801.
4. Baluška, F., & Reber, A. (2019). Sentience and consciousness in single cells: how the first minds emerged in unicellular species. BioEssays, 41(3), 1800229.
5. Thompson, E. (2022). Could All Life Be Sentient? Journal of Consciousness Studies, 29(3-4), 229-265.
6. Ball, P. (2022). The Book of Minds: How to Understand Ourselves and Other Beings, from Animals to AI to Aliens. University of Chicago Press.
7. Chalmers, D. J. (2023). Could a large language model be conscious? arXiv preprint arXiv:2303.07103.
8. y Arcas, B. A. (2022). Do large language models understand us?. *Daedalus*, *151*(2), 183-197.
9. VanRullen, R. and Kanai, R. (2021) Deep learning and the Global Workspace Theory. *Trends in Neurosciences* 44, 692–704
10. Juliani, A., Arulkumaran, K., Sasai, S., & Kanai, R. (2022). On the link between conscious function and general intelligence in humans and machines. *Transactions on Machine Learning Research*.
11. Rumelhart, D.E., ed. (1999) *Parallel distributed processing. 1: Foundations / David E. Rumelhart*, (12. print.), MIT Press.
12. Aru, J. *et al.* (2020) Cellular Mechanisms of Conscious Processing. *Trends in Cognitive Sciences* 24, 814–825
13. Shine, J.M. (2021) The thalamus integrates the macrosystems of the brain to facilitate complex, adaptive brain network dynamics. *Progress in Neurobiology* 199, 101951
14. Llinás, R. and Ribary, U. (2001) Consciousness and the brain. The thalamocortical dialogue in health and disease. *Annals of the New York Academy of Sciences* 929, 166–175
15. Tasserie, J. *et al.* (2022) Deep brain stimulation of the thalamus restores signatures of consciousness in a nonhuman primate model. *Sci. Adv.* 8, eabl5547

16. Koch, C. *et al.* (2016) Neural correlates of consciousness: progress and problems. *Nat Rev Neurosci* 17, 307–321

17. Aru, J. *et al.* (2019) Coupling the State and Contents of Consciousness. *Front. Syst. Neurosci.* 13, 43

18. Parvizi, J. and Damasio, A. (2001) Consciousness and the brainstem. *Cognition* 79, 135–160

19. Shine, J.M. (2023) Neuromodulatory control of complex adaptive dynamics in the brain. Interface focus 13, 20220079

20. Snider, S. B., Bodien, Y. G., Bianciardi, M., Brown, E. N., Wu, O., & Edlow, B. L. (2019). Disruption of the ascending arousal network in acute traumatic disorders of consciousness. *Neurology*, *93*(13), e1281-e1287.

21. Edlow, B. L., Takahashi, E., Wu, O., Benner, T., Dai, G., Bu, L., ... & Folkerth, R. D. (2012). Neuroanatomic connectivity of the human ascending arousal system critical to consciousness and its disorders. *Journal of Neuropathology & Experimental Neurology*, *71*(6), 531-546.

22. Spindler, L. R., Luppi, A. I., Adapa, R. M., Craig, M. M., Coppola, P., Peattie, A. R., ... & Stamatakis, E. A. (2021). Dopaminergic brainstem disconnection is common to pharmacological and pathological consciousness perturbation. *Proceedings of the National Academy of Sciences*, *118*(30), e2026289118.

23. Hindman, J., Bowren, M. D., Bruss, J., Wright, B., Geerling, J. C., & Boes, A. D. (2018). Thalamic strokes that severely impair arousal extend into the brainstem. *Annals of neurology*, *84*(6), 926-930.

24. Merker, B. (2007) Consciousness without a cerebral cortex: A challenge for neuroscience and medicine. *Behav Brain Sci* 30, 63–81

25. Shine, J.M. (2022) Adaptively navigating affordance landscapes: How interactions between the superior colliculus and thalamus coordinate complex, adaptive behaviour. *Neuroscience & Biobehavioral Reviews* 143, 104921

26. Barron, A.B. and Klein, C. (2016) What insects can tell us about the origins of consciousness. *Proc Natl Acad Sci USA* 113, 4900–4908

27. Mitchell, M. and Krakauer, D.C. (2023) The debate over understanding in AI's large language models. *Proc. Natl. Acad. Sci. U.S.A.* 120, e2215907120

28. Block, N. (1995) On a confusion about a function of consciousness. *Behav Brain Sci* 18, 227–247

29. Uexküll, J. von (2010) *A foray into the worlds of animals and humans: with A theory of meaning*, (1st University of Minnesota Press ed.), University of Minnesota Press.

30. Wakakuwa, M. *et al.* (2007) Spectral Organization of Ommatidia in Flower-visiting Insects. *Photochemistry and Photobiology* 83, 27–34

31. Chen, Q. *et al.* (2012) Reduced Performance of Prey Targeting in Pit Vipers with Contralaterally Occluded Infrared and Visual Senses. *PLoS ONE* 7, e34989

32. Gibson, J.J. (1966) *The Senses Considered as Perceptual Systems.*, Houghton Mifflin, Boston.

33. Greeno, J.G. (1994) Gibson's affordances. *Psychological Review* 101, 336–342

34. Pezzulo, G. and Cisek, P. (2016) Navigating the Affordance Landscape: Feedback Control as a Process Model of Behavior and Cognition. *Trends in Cognitive Sciences* 20, 414–424

35. Vaswani, A. *et al.* (2017) Attention Is All You Need. DOI: 10.48550/ARXIV.1706.03762

36. Brown, T.B. *et al.* (2020) Language Models are Few-Shot Learners. DOI: 10.48550/ARXIV.2005.14165

37. Nagel, T. (1974) What Is It Like to Be a Bat? *The Philosophical Review* 83, 435

38. Radford, A., Kim, J. W., Hallacy, C., Ramesh, A., Goh, G., Agarwal, S., ... & Sutskever, I. (2021). Learning transferable visual models from natural language supervision. In International conference on machine learning (pp. 8748-8763).

39. Driess, D., Xia, F., Sajjadi, M. S., Lynch, C., Chowdhery, A., Ichter, B., ... & Florence, P. (2023). Palm-e: An embodied multimodal language model. arXiv preprint arXiv:2303.03378.

40. Roli, A., Jaeger, J., & Kauffman, S. A. (2022). How organisms come to know the world: fundamental limits on artificial general intelligence. *Frontiers in Ecology and Evolution*, *9*, 1035.

41. Fultot, M., & Turvey, M. T. (2019). Von Uexküll's theory of meaning and Gibson's organism–environment reciprocity. *Ecological Psychology*, *31*(4), 289-315.

42. Damasio, A., & Damasio, H. (2022). Homeostatic feelings and the biology of consciousness. *Brain*, *145*(7), 2231-2235.

43. Damasio, A., & Damasio, H. (2022). Feelings are the source of consciousness. *Neural Computation*, 1-10.

44. Damasio, A. (2021). *Feeling & knowing: Making minds conscious*. Pantheon.

45. Panksepp, J. (2005). Affective consciousness: Core emotional feelings in animals and humans. *Consciousness and cognition*, *14*(1), 30-80.

46. Solms, M. (2021). *The hidden spring: A journey to the source of consciousness*. Profile books.

47. Seth, A.K. and Bayne, T. (2022) Theories of consciousness. Nat Rev Neurosci 23, 439–452

48. Dehaene, S., & Changeux, J. P. (2011). Experimental and theoretical approaches to conscious processing. Neuron, 70(2), 200-227.

49. Mashour, G.A. *et al.* (2020) Conscious Processing and the Global Neuronal Workspace Hypothesis. *Neuron* 105, 776–798

50. Tononi, G. (2008). Consciousness as integrated information: a provisional manifesto. *The Biological Bulletin*, *215*(3), 216-242

51. Tononi, G. *et al.* (2016) Integrated information theory: from consciousness to its physical substrate. *Nat Rev Neurosci* 17, 450–461

52. Redinbaugh, M.J. *et al.* (2020) Thalamus Modulates Consciousness via Layer-Specific Control of Cortex. *Neuron* 106, 66-75.e12

53. Bachmann, T., Suzuki, M., & Aru, J. (2020). Dendritic integration theory: a thalamo-cortical theory of state and content of consciousness. Philosophy and the Mind Sciences, 1(II).

54. Bell, P.T. and Shine, J.M. (2016) Subcortical contributions to large-scale network communication. *Neuroscience & Biobehavioral Reviews* 71, 313–322

55. Shine, J. M., Lewis, L. D., Garrett, D. D., & Hwang, K. (2023). The impact of the human thalamus on brain-wide information processing. *Nature Reviews Neuroscience*, 1-15

56. Suzuki, M., Pennartz., C & Aru, J (2023). How deep is the brain? The shallow brain hypothesis. *Nature Reviews Neuroscience*, 1-15.

57. Supèr, H., Spekreijse, H., & Lamme, V. A. (2001). Two distinct modes of sensory processing observed in monkey primary visual cortex (V1). Nature neuroscience, 4(3), 304-310.

58. Lamme, V. A. (2006). Towards a true neural stance on consciousness. Trends in cognitive sciences, 10(11), 494-501.

59. Imas, O. A., Ropella, K. M., Ward, B. D., Wood, J. D., & Hudetz, A. G. (2005). Volatile anesthetics disrupt frontal-posterior recurrent information transfer at gamma frequencies in rat. Neuroscience letters, 387(3), 145-150.

60. Boly, M., Garrido, M. I., Gosseries, O., Bruno, M. A., Boveroux, P., Schnakers, C., ... & Friston, K. (2011). Preserved feedforward but impaired top-down processes in the vegetative state. Science, 332(6031), 858-862.

61. Ku, S. W., Lee, U., Noh, G. J., Jun, I. G., & Mashour, G. A. (2011). Preferential inhibition of frontal-to-parietal feedback connectivity is a neurophysiologic correlate of general anesthesia in surgical patients. PloS one, 6(10), e25155.
62. Bayne, T. (2012). *The unity of consciousness*. Oxford University Press.
63. Singer, W. (1998). Consciousness and the structure of neuronal representations. *Philosophical transactions of the royal society of London. Series B: biological sciences*, *353*(1377), 1829-1840.
64. Singer, W. (2001). Consciousness and the binding problem. *Annals of the New York Academy of Sciences*, *929*(1), 123-146.
65. Larkum, M. (2013). A cellular mechanism for cortical associations: an organizing principle for the cerebral cortex. *Trends in neurosciences*, *36*(3), 141-151.
66. Goyal, A., Didolkar, A., Lamb, A., Badola, K., Ke, N. R., Rahaman, N., ... & Bengio, Y. (2021). Coordination among neural modules through a shared global workspace. *arXiv preprint arXiv:2103.01197*.
67. Juliani, A., Kanai, R., & Sasai, S. S. (2022). The perceiver architecture is a functional global workspace. In *Proceedings of the Annual Meeting of the Cognitive Science Society* (Vol. 44, No. 44).
68. Koch, C. (2019). The feeling of life itself: why consciousness is widespread but can't be computed. Mit Press.
69. Tononi, G., & Koch, C. (2015). Consciousness: here, there and everywhere?. *Philosophical Transactions of the Royal Society B: Biological Sciences*, *370*(1668), 20140167.
70. Krom, A. J., Marmelshtein, A., Gelbard-Sagiv, H., Tankus, A., Hayat, H., Hayat, D., ... & Nir, Y. (2020). Anesthesia-induced loss of consciousness disrupts auditory responses beyond primary cortex. *Proceedings of the National Academy of Sciences*, *117*(21), 11770-11780.
71. Hayat, H., Marmelshtein, A., Krom, A. J., Sela, Y., Tankus, A., Strauss, I., ... & Nir, Y. (2022). Reduced neural feedback signaling despite robust neuron and gamma auditory responses during human sleep. *Nature neuroscience*, *25*(7), 935-943.
72. Suzuki, M., & Larkum, M. E. (2020). General anesthesia decouples cortical pyramidal neurons. *Cell*, *180*(4), 666-676.
73. Man, K. and Damasio, A. (2019) Homeostasis and soft robotics in the design of feeling machines. *Nat Mach Intell* 1, 446–452
74. Seth, A. (2021). *Being you: A new science of consciousness*. Penguin.
75. Thompson, E. (2010) *Mind in life: biology, phenomenology, and the sciences of mind*, (First Harvard University Press paperback edition.), The Belknap Press of Harvard University Press
76. Seth, A.K. and Tsakiris, M. (2018) Being a Beast Machine: The Somatic Basis of Selfhood. *Trends in Cognitive Sciences* 22, 969–981

77. Deacon, T.W. (2012) *Incomplete nature: how mind emerged from matter*, (1st ed.), W.W. Norton & Co
78. Louie, A. H. (2013). *More than life itself: A synthetic continuation in relational biology* (Vol. 1). Walter de Gruyter.
79. Rosen, R. (1991). Life itself: a comprehensive inquiry into the nature, origin, and fabrication of life. Columbia University Press.
80. Rosen, R. (2000). Essays on life itself. Columbia University Press.
81. Moreno, A., & Mossio, M. (2015) Biological Autonomy A Philosophical and Theoretical Enquiry. Springer.
82. Nicholson, D. J. (2019). Is the cell really a machine?. *Journal of theoretical biology*, *477*, 108-126.
83. Kandel, E. R., Schwartz, J. H., Jessell, T. M., Siegelbaum, S., Hudspeth, A. J., & Mack, S. (Eds.). (2000). Principles of neural science (Vol. 4, pp. 1227-1246). New York: McGraw-hill.
84. Lane, N. (2022). Transformer: The Deep Chemistry of Life and Death. Profile Books.
85. Searle, J. R. (1992). The rediscovery of the mind. MIT press.
86. Penrose, R. (1994). Shadows of the Mind. Oxford: Oxford University Press.
87. Gidon, A., Aru, J., & Larkum, M. E. (2022). Does brain activity cause consciousness? A thought experiment. PLoS Biology, 20(6), e3001651.
88. Butlin, P. et al. (2023). Consciousness in Artificial Intelligence: Insights from the Science of Consciousness. Arxiv preprint.
89. Richards, B. A., Lillicrap, T. P., Beaudoin, P., Bengio, Y., Bogacz, R., Christensen, A., ... & Kording, K. P. (2019). A deep learning framework for neuroscience. *Nature neuroscience*, *22*(11), 1761-1770.
90. Zador, A., Escola, S., Richards, B., Ölveczky, B., Bengio, Y., Boahen, K., ... & Tsao, D. (2023). Catalyzing next-generation artificial intelligence through neuroai. *Nature communications*, *14*(1), 1597.
91. Doerig, A., Sommers, R. P., Seeliger, K., Richards, B., Ismael, J., Lindsay, G. W., ... & Kietzmann, T. C. (2023). The neuroconnectionist research programme. *Nature Reviews Neuroscience*, 1-20.
92. Metzinger, T. (2021) Artificial Suffering: An Argument for a Global Moratorium on Synthetic Phenomenology. *J. AI. Consci.* 08, 43–66.
93. Taleb, N.N. (2018) *Skin in the game: hidden asymmetries in daily life*, (First edition.), Random House

94. Jonas, H. (2001) *The phenomenon of life: toward a philosophical biology*, Northwestern University Press